\documentclass[12pt]{article}
\usepackage[utf8]{inputenc}
\usepackage[dvipsnames]{xcolor}
\usepackage[round]{natbib}
\usepackage{tikz}
\usepackage[colorlinks=false,urlcolor=blue,citecolor=magenta]{hyperref}
\usepackage{amssymb,amsmath,amsfonts}
\usepackage{epsfig}
\usepackage{epsfig}
\usepackage{graphicx}
\usepackage{epstopdf}
\usepackage{bbold}
\usepackage{caption}
\usepackage{subcaption}
\usepackage{amscd}
\usepackage{amsthm}
\usepackage{latexsym}
\usepackage{float}
\usepackage{mathtools}
\usepackage{amsbsy}
\usepackage{bbm}
\usepackage[english]{babel}
\usepackage{psfrag}
\usepackage{tabularx}
\usepackage{colortbl}
\usepackage{braket}
\usepackage{csquotes}
\usepackage[stable]{footmisc}
\usepackage{setspace}
\usepackage{authblk}
\usepackage{lmodern}

\usepackage{comment}

\newcommand{\be}{\begin{eqnarray}}
\newcommand{\ee}{\end{eqnarray}}
\newcommand{\beq}{\begin{eqnarray}}
\newcommand{\eeq}{\end{eqnarray}}

\title{\bf Holographic Strange Metals for Philosophers and Physicists}
\author[1]{Enrico Cinti}
\author[2]{Sebastian De Haro}
\author[3]{Mark S.~Golden}
\author[4]{Umut G\"ursoy\footnote{{\it We dedicate this paper to the memory of Umut G\"ursoy,} who tragically passed away on 24 April 2025, during the last stages of the paper's completion.}}
\author[5]{Henk T.~C.~Stoof}
\affil[1]{Department of Philosophy, University of Geneva}
\affil[1,2]{Institute for Logic, Language and Computation, University of Amsterdam}
\affil[2,3]{Institute of Physics, University of Amsterdam}
\affil[3]{Dutch Institute for Emergent Phenomena, University of Amsterdam}
\affil[4,5]{Institute for Theoretical Physics and Centre for Extreme Matter and Emergent Phenomena, Utrecht University}
\date{\today}

\begin{document}
\maketitle

\begin{abstract}

This paper introduces the physics and philosophy of strange metals, which are characterized by unusual electrical and thermal properties that deviate from conventional metallic behaviour. The anomalous strange-metal behaviour discussed here appears in the normal state of a copper-oxide high-temperature superconductor, and it cannot be described using standard condensed-matter physics. Currently, it can only be described through a holographic dual, viz.~a four-dimensional black hole in anti-de Sitter spacetime. This paper first introduces the theory of, and specific experiments carried out on, strange metals. Then it discusses a number of philosophical questions that strange metals open up regarding the experimental evidence for holography and its realist interpretation. Strange metals invert the explanatory arrows, in that usual holographic arguments are seen as giving explanations of the bulk quantum-gravity theory from the boundary. By contrast, the aim here is, by using holography, to explain the experimentally discovered and anomalous properties of strange metals.

\end{abstract}


\tableofcontents

\newpage

\section{Introduction}\label{intro}

Condensed-matter physics has proven to be an important and rich avenue for advances in experimental and theoretical physics. An important class of systems studied in condensed-matter physics are strange metals. Although strange metals have been studied experimentally for several decades, their only known theoretical description is through a version of holography in AdS-CFT, called ``semi-holography''. In short, semi-holography is a hybrid version of holography where a weakly-coupled field---here, a fundamental fermion---is coupled to a strongly-coupled composite operator in a CFT that has a holographic dual. (For a more detailed description, see Section \ref{adscft}.)

This connection between condensed-matter physics and AdS-CFT is interesting for a variety of reasons. As a theory of quantum gravity, AdS-CFT would itself benefit from the empirical support and precise testing that one envisages could come from strange metals. Likewise, the philosophical study of holography could advance by going beyond its string-theoretic realizations and, through experimentation, establish contact with the real world in a true ``experimental philosophy'', akin to what we have seen for quantum-information theory. Philosophy can bring clarity to issues that are important for both experimentalists and theorists, like the emergence and explanatory value of the strange-metal phase of a material. For these reasons, it is natural to say that strange metals and their holographic description lie at the interface of theory, experiment and philosophy.

Because of this close intertwining between theory, experiment and philosophy in a set of mutual dependency relations, where the current limitations of each approach can, to some extent, be mitigated or, at least, eased by the virtues of the two others, we will argue that strange metals are a privileged locus for doing natural philosophy. In this paper, we aim to initiate such a bootstrapping approach to strange metals and their holography. Thus, after reviewing the basic theoretical and experimental features of holographic strange metals, we highlight a number of philosophical questions that they raise and some avenues for how to resolve them.

In short, a strange metal is a metal whose behaviour is not described by the standard Fermi-liquid theory that is used to explain the phenomenology of most known metals. One important class of strange metals that we will focus on in this paper is the cuprates, or copper-oxide superconducting materials that contain layers of copper and oxygen atoms. They are high-temperature superconductors, and the strange-metal phase is the phase above the critical temperature, where the superconductivity is lost. 

The anomalous behaviour of a strange metal includes the linear temperature dependence of the resistivity, which for the optimal doping level (with the highest superconducting critical temperature) spans from just above the critical temperature up to the material's melting point. Thus, in contrast to normal metals, no quadratic ($T^2$) dependence of the resistivity on the temperature is seen at low temperatures, as well as no saturation in the resistivity at high temperatures. Other anomalous features include a momentum-dependence of the exponent describing the power-law behaviour of the self-energy of the electrons, observed using precision angle-resolved photoemission spectroscopy (ARPES) experiments, which are a sophisticated version of the photoelectric effect.

From the point of view of physics, strange metals are interesting both because of these unexplained anomalous behaviours and because it is believed that understanding the strange-metal phase of a material is crucial for understanding high-temperature superconductivity. Moreover, modelling strange metals using holographic and semi-holographic tools prompts interesting questions about the conceptual role of semi-holography in this kind of theoretical and experimental work: in particular, whether its use is explanatory in a substantive sense, or whether it is a mere computational device, and whether the gravitational dual to a strange metal can be said to emerge. 

Furthermore, holographic strange metals raise the intriguing possibility that these experiments might provide concrete experimental evidence for the physical significance of a holographic description. This means that the experiments could provide evidence for the behaviour of a physical system whose modelling depends crucially (perhaps even essentially) on a holographic description. 

Such claims about the role of semi-holography in experimental confirmation invite careful conceptual investigation, which we begin in this paper to undertake. As a starting point for the ``bootstrapping'' we discussed above, we have chosen four philosophical topics, and we address the corresponding questions: emergence, explanation, scientific realism and the distinction between semi-holography and holography. Emergence has received significant attention both in the physics and philosophy literature on dualities, and holography in particular: thus we can adapt that analysis to cast light on strange metals. Does the strange-metal phase emerge, or does the black-hole description emerge from the strange metal? Explanation and scientific realism allow us to explore the possibility of using evidence from holographic strange metals to support holography: Does holography provide explanations for strange metals? And, given the available evidence, should a scientific realist be committed to the holographic features of strange metals? Finally, the distinction between semi-holography and holography is a foundational question for holographic strange metals, since it aims to clarify the conceptual framework behind the application of holography to strange metals.

By engaging with the experiments, the theory and the philosophy of strange metals and semi-holography, we aim to set the stage for a fruitful discussion and to give philosophers of physics and philosophically-minded physicists the tools to explore the foundations of holographic strange metals.\footnote{Note that there is a rich philosophical literature on dualities, and holography in particular. Relevant works are: Rickles (2011), Matsubara (2013), Teh (2013), Dieks et al.~(2015), Dawid (2017), De Haro (2017), Le Bihan and Read (2018) and De Haro and Butterfield (2025). However, most of this literature does not discuss the question of the empirical evidence for holography. To the best of our knowledge, the only paper that directly addresses this topic is Dardashti et al.~(2018), which focusses on cases that are very different from the one we discuss in this paper.} 
We take philosophy of physics to be central to this endeavour, because the questions that we are facing combine technical and conceptual aspects: answering them requires both careful attention to the technical detail of the physical description of strange metals, and conceptual rigour in understanding 
these models. 

This paper is structured as follows: Section \ref{theory} gives an overview of the basic theoretical underpinnings of holographic strange metals and reviews the experimental evidence for them. Section \ref{philosophy} discusses the philosophical issues that they raise and how to approach them. Section \ref{conclusion} concludes.

\section{Theory and Experiment}\label{theory}

This Section introduces the theoretical and experimental aspects of strange metals. Section \ref{condensed} first discusses the difficulties of describing strange metal within the conventional framework of condensed-matter physics. Section \ref{adscft} introduces the idea of ``semi-holography'' as a variation of AdS-CFT, and Section \ref{semistrange} discusses semi-holography as an alternative framework where strange metals can be modelled successfully. Finally, Section \ref{experiment} reviews the current status of strange-metal experiments.

\subsection{Condensed-Matter Physics}\label{condensed}

The study of metals has played a major role in condensed-matter physics, leading ultimately to the ``Standard Theory of Metals" that is based on the famous Fermi-liquid theory of Landau and explains the phenomenology of many known metals. However, nowadays we still find ``strange" metals that elude our understanding and require the development of new physical theories. To understand the reasons behind their name we have to define what we mean by ``normal" metals. In a simplified picture, a metal consists of a lattice of fixed atoms with conduction electrons that are loosely bound to these atoms and can readily hop from one lattice site to the nearest neighbour (or next-nearest neighbour, etc.). Importantly, these electrons are fermions subject to the Pauli principle that states that there cannot be two electrons in the same quantum state. Due to this restriction and neglecting interactions, the electrons thus form a liquid with quantum states filled up to a Fermi surface, and the low-energy excitations correspond to adding a particle to or removing a particle from the Fermi sea just above or just below the Fermi surface. The crucial insight of Landau, made in the context of a different system of fermions, namely liquid $^3$He, was that the interactions between the electrons, irrespective of the strength of the interactions at the lattice (ultraviolet) scale, often still allow for a low-energy (infrared) description in terms of fermionic particles and holes, although they are not the original particles of the non-interacting theory. Indeed, these emergent particle-like excitations have different effective properties than electrons of the standard model of particle physics, possessing for example a different mass and magnetic moment, and are consequently referred to as quasiparticles. The resulting Fermi-liquid framework then provides a weakly interacting picture to study the physical properties of metals, even when the interactions between the electrons are moderately strong. This led to an understanding of many metallic properties such as, for instance, the specific heat, the magnetic susceptibility, and the $T^2$ dependence of the resistivity at low temperatures $T$. Building on Fermi-liquid theory, the Bardeen-Cooper-Schrieffer (BCS) theory of superconductivity then explains why common metals become superconductors when cooled sufficiently close to absolute zero. \\
\\
{\it Strange metals as the normal state phase from which high-$T_c$ superconductors are born.} In 1986, an experiment on a copper-oxide compound brought to light the phenomenon of high-temperature superconductivity, sparking a new experimental and theoretical interest in this class of materials referred to as the ``cuprates''. As the precise mechanism causing their high superconducting critical temperatures is still unclear, there is hope for the possibility of achieving even room-temperature superconductivity, which makes this problem of particular practical appeal. The strange metal of interest to us here is the normal, i.e.~non-superconducting, phase of these high-$T_c$ superconductors and it is an example of a non-Fermi-liquid metal, showing a very different behaviour compared to that of normal metals that fall under Landau's successful Fermi-liquid framework. Perhaps the most famously strange property of this phase is the anomalous behaviour of the resistivity, which is exactly linear in $T$ from the lowest temperatures reached just above the critical temperature for superconductivity to the highest possible temperatures before the material melts. This linear resistivity is a universal feature of all strange metals. For this reason we focus our paper on the study of holographic models dual to field theories which deliver such a linear behaviour, although there are other anomalous features that characterize the strange-metal phase in the cuprates, such as the doping and temperature dependence of the Hall number and the Hall angle.      

\subsection{Semi-Holography and AdS-CFT}\label{adscft}

\begin{figure}
\begin{center}
\includegraphics[height=6cm]{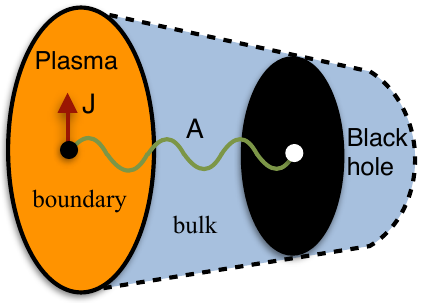}
\caption{\small Sketch of the holographic correspondence, from (G\"ursoy, 2021).}
\label{Holo1}
\end{center}
\end{figure}

The AdS-CFT correspondence gives a concrete realization of the holographic principle relating field theory to gravity in one higher dimension (for short, we will simply speak about ``holography''). Holography is regarded as a standard tool for modelling strongly correlated quantum systems, not only in high-energy (Casalderrey-Solana et al. 2014) but also in condensed-matter physics (see Hartnoll et al.~2018, Zaanen et al.~2015 for reviews). Briefly, the correspondence---in the form that will be relevant for us here---relates Einstein’s theory of gravity in the bulk of a $d+1$-dimensional\footnote{There may be additional ``internal'' dimensions in the bulk.} spacetime with a black hole at the centre, see Figure \ref{Holo1}, to a strongly coupled, non-gravitational theory of a quantum thermal state on the $d$-dimensional boundary of this spacetime (Maldacena, 1998; Gubser et al 1998; Witten 1998).\footnote{For a review of gauge-gravity dualities oriented towards a philosophical audience, see De Haro et al.~(2016).} 
It maps the collective transport — denoted by the current $J$ at the black point in Figure \ref{Holo1} (G\"ursoy, 2021) — to fluctuations near the horizon — at the white point — through propagation of a bulk wave $A$ toward the black hole. Universal properties of horizon geometry then lead to constraints on transport, see e.g. (Policastro et al. 2001; Kovtun et al. 2005). 

The original formulation of holography maps the computation of the correlation function of {\em composite operators} such as the current $J$ in Figure \ref{Holo1} on the boundary to the dynamics of a corresponding bulk field $A$ in the bulk.   However, this original formulation does not allow for the calculation of single-particle correlation functions. For instance the single-electron Green's function $G(\omega,  {\bf k})$ with $\omega$ and  {\bf k} denoting the frequency and momentum respectively, is a fundamental observable, directly measured by angle-resolved photoemission spectroscopy (ARPES) experiments (see Section \ref{experiment}), that cannot be obtained using the original formulation of the correspondence. A fundamental difference between correlation functions of single-particle and composite operators is that the former satisfies the so-called ``sum-rule'', i.e. the integral over frequency of the imaginary part of $G(\omega,  {\bf k})$ is unity, while the latter does not (G\"ursoy et al. 2012).  

\begin{figure}
\begin{center}
\includegraphics[height=7cm]{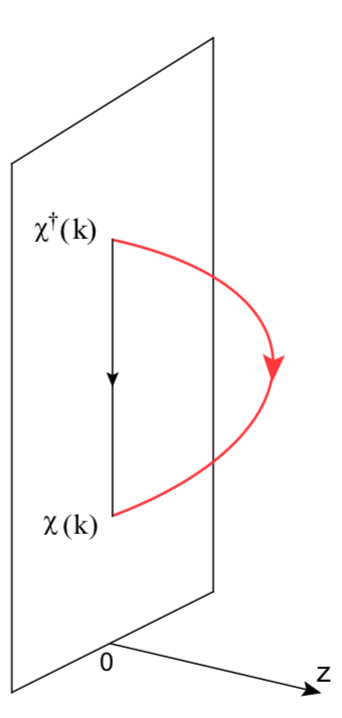}
\caption{\small Two channels of propagation for the boundary elementary field $\chi$ in semi-holography: on the boundary, and through the bulk.} 
\label{Semi}
\end{center}
\end{figure}

{\em Semi-holography} was first introduced in (Faulkner and Polchinski 2011) to remedy this situation. The idea here is to couple a canonical field $\chi$ to a composite operator ${\cal O}$ in the strongly coupled quantum theory - which is usually taken to be a conformal field theory (CFT) - by introducing a coupling proportional to $ g \chi \cal{O}^\dagger$ in the action, where $g$ is a coupling constant. In addition, one assumes that the elementary field $\chi$ has a standard kinetic term. This allows for two separate channels for propagation of $\chi$ on the boundary quantum theory: (i) it can propagate via the standard kinetic term which leads to the standard Green's function of the ``would-be'' free $\chi$ field $G_{\textrm{free}}(\omega,  {\bf k})$ , or (ii) it can turn into ${\cal O}$ which propagates through the CFT correlator $\langle {\cal O}^\dagger {\cal O} \rangle$ and turns back into $\chi$. Clearly, the second channel can repeat multiple times and summing over all number of repetitions produce a Green's function of the form $G(\omega,  {\bf k}) = 1/[G_{\textrm{free}}(\omega,  {\bf k}) + |g|^2 \langle {\cal O}^\dagger {\cal O} \rangle]$. In this way one obtains the standard form of the Green's function for an {\em interacting } field $\chi$ with the self-energy term given by $\Sigma(\omega,  {\bf k}) \equiv |g|^2  \langle {\cal O}^\dagger {\cal O} \rangle)$. One maps {\em only} this self-energy part, $\Sigma$, holographically to the bulk gravity theory. That is, one computes $\Sigma$ by considering propagation of a bulk field $\Psi$ dual to the operator ${\cal O}$ in the black-hole background, hence the name {\em semi}-holography. One can therefore view the two aforementioned modes of propagation of the $\chi$ field as (i) on the boundary and (ii) in the bulk with the vertex representing coupling of $\chi$ with the operator ${\cal O}$; see Figure \ref{Semi}. 

As we will discuss in more detail in the next section, because of this hybrid construction semi-holography cannot, as such, be viewed as holography. However, as was first understood in (G\"ursoy et al. 2012), and as we review below, this undesired feature can be side-stepped by realizing that the same conclusion can be achieved {\em within} a holographic framework, by allowing for alternative boundary conditions for the bulk fields. To illustrate this, consider the standard holographic dictionary in which we view the boundary value of the bulk field $\Psi$ as the source for the corresponding boundary operator ${\cal O}$. Now instead of considering $\chi$ as an additional elementary field on the boundary, we can identify it with the source of the operator ${\cal O}$, hence $\chi = \lim_{z\to 0} \Psi$ where the boundary is located at $z=0$\footnote{For technical reasons one usually takes this at a finite but small value.}. It turns out, see (G\"ursoy et al. 2012) and (Contino and Pomarol 2004), that one can define the variational problem of the bulk action in a way that is consistent with the addition of an arbitrary boundary action that depends on this source $\chi$ which we can choose to be the kinetic term for $\chi$. This makes the source $\chi$ dynamical. Therefore, instead of fixing it, we take the path integral over this dynamical source. This then, immediately leads to the same Green's function $G(\omega,  {\bf k}) = 1/[G_{\textrm{free}}(\omega,  {\bf k}) + |g|^2 \langle {\cal O}^\dagger {\cal O} \rangle ]$. This corresponds to an alternative quantization of the AdS bulk theory, where - instead of fixing the boundary value of the bulk field $\Psi$ - one integrates over it, fixing instead another source which is the Legendre conjugate of $\chi$. In this way, we can subsume semi-holography into a holographic framework with an alternative choice of boundary conditions: we {\it reduce semi-holography to holography.}

\subsection{Semi-Holography and Strange Metals}\label{semistrange}

In the context of bottom-up model building, the question arises regarding the ingredients required for a semi-holographic description of strange metals. As the strange-metal phase occupies a large area in the doping and temperature-dependent phase diagram of the cuprates, its thermodynamics is described by a chemical potential $\mu$ and a temperature $T$. Using the holographic dictionary, the spacetime background of the Dirac equation requires that the theory of gravity is coupled to electromagnetism, i.e.~the Einstein-Maxwell theory. \footnote{The argument is as follows: boundary temperature corresponds holographically to black hole temperature, and a chemical potential for a conserved U(1) charge on the boundary corresponds to a bulk U(1) gauge field. The minimal, two‑derivative effective bulk theory that contains these two ingredients is the Einstein-Maxwell gravity theory.} 
As we discuss below, the resulting theory has one important feature in common with strange metals, but nevertheless it is not an appropriate theory for a strange metal. This is because a defining property of the strange metal is its linear temperature dependence of the resistivity. To be able to reproduce this we need the black-hole entropy to also be linear in temperature. Physically this can be qualitatively understood from the fact that in holography the viscosity of the dual quantum field theory is proportional to the entropy density, and since the resistivity is proportional to the viscosity we need a black-hole entropy proportional to $T$. The black-hole background in Einstein-Maxwell theory is well known, and is known as the Reissner-Nordstr\"om solution. Its entropy is not linear in the temperature, and in fact it is non-zero at zero temperature, which implies that this solution is unstable and a phase transition will occur at low temperatures to another black-hole solution. 

To remedy this situation, we add an additional scalar field to the model, which results in the Einstein-Maxwell-Dilaton theory. A class of black-hole solutions exists in this theory, which \textit{are} characterized by a dynamical critical exponent $z$ and a hyperscaling violating exponent $\theta$. The black-hole entropy is then linear in the temperature if the condition $\theta/z=-1$ is satisfied. This will reproduce the linear-in-$T$ resistivity of the strange metal as we have seen. However, there is one more piece of experimental information that needs to be incorporated for an appropriate model of a strange metal. Angle-resolved photoemission spectroscopy (ARPES) measurements of strange-metal cuprates (Smit et al. 2024) along the so-called nodal k-space direction (along the tetragonal Brillouin diagonal) have shown that the imaginary part of the electron self-energy very accurately conforms to the ``power-law-liquid" result $\hbar \Sigma''(\omega, \vec{k}) \propto \omega^{2\alpha}$, where $\alpha$ is doping dependent. Crucially, these recent ARPES data (Smit et al. 2024) have also shown that $\alpha$ is also $k$-dependent: $\alpha(k) = \alpha [1 - ({k}-k_{F})/k_{F} ]$. At optimal doping (maximal superconducting critical temperature), $\alpha = 1/2$ and then the observed frequency and temperature dependence of the imaginary part of the self-energy is - bar the $k$-dependence - equal to that of the famous marginal Fermi liquid. 
Consequently, at the Fermi energy, the power-law exponent $\alpha$ becomes momentum independent and obeys power-law scaling with a critical exponent that depends on the strange metal's doping level. This fully conforms to the picture of the physics of the strange metal being governed by a quantum-critical phase, in contrast to a quantum-critical point, which, as we will argue next, makes it ideal to model with holography.

To incorporate this last piece of experimental information while having the electron self-energy dominated by its frequency dependence, we consider the limit $z \rightarrow \infty$, in which case semi-holography will give a self-energy of the form $\hbar \Sigma(\omega, \vec{k}) \propto \omega (-\omega^2)^{\nu_{\vec k} -1/2}$, with the momentum dependence only entering via the exponent $\nu_{\vec k}$. It should be noted that the Reissner-Nordstr\"om black hole also leads to a self-energy of this form with the dynamical exponent $z = \infty$, but unfortunately without having a linear-in-$T$ resistivity. Within the Einstein-Maxwell-Dilaton theory, all requirements to correctly model a strange metal can be fulfilled by considering the simultaneous limits $\theta \rightarrow -\infty$ and $z \rightarrow \infty$, with $\theta/z=-1$. In this context, the Gubser-Rocha model plays a special role as it gives a fully analytic black-hole solution exactly valid in this combination of limits, leading to $\nu_{\vec k} = 2q\hbar v_F|\vec{k}|/\mu$, where $v_F$ is the Fermi velocity (Gubser and Roche 2010). Note here especially the proportionality of the exponent $\nu_{\vec k}$ to the dimensionless charge of the Dirac fermion $q$, which makes it ideal to describe the strange metal as a critical phase by simply making $q$ doping dependent such that $\nu_{\vec k} = \alpha |\vec{k}|/k_F$, with $\hbar k_F$ the momentum of the Fermi surface (Mauri et al. 2024).     
 
\subsection{The ARPES Experiments}\label{experiment}

In this Section, we discuss how the self-energy of the electrons in strange metals can be probed experimentally. In the context of this paper, it is of particular interest to use experiments to check whether or not the self-energy displays the expected frequency dependence that is computed via holographic methods. However, we do mention here that the experimentalist team behind (Smit et al. 2024) first observed the tell-tale signs of the $k$-dependence of the self-energy in the ARPES data without prior knowledge of the holographic predictions.    


The experimental tool providing access to the ($\omega,T$) dependence of the self-energy is Angle-Resolved Photoelectron Spectroscopy (ARPES).\footnote{For a recent review of ARPES methods, see Zhang et al. 2022.} This technique employs the photoelectric effect to measure the properties of the electrons in a strange metal. A clean surface of the sample under investigation is exposed to a photon beam of sufficient energy, and electron emission from the surface follows on absorption of a photon. The resulting photoelectrons are analyzed using a spectrometer that tracks both their emission angle $\theta$ with respect to the sample normal and their kinetic energy $E_{\mbox{\tiny{kin}}}$. From these quantities, the energy and wave vector $\textbf{k}$ of the electrons in the occupied part of the band structure can be extracted.

\begin{figure}
\begin{center}
\includegraphics[height=6cm]{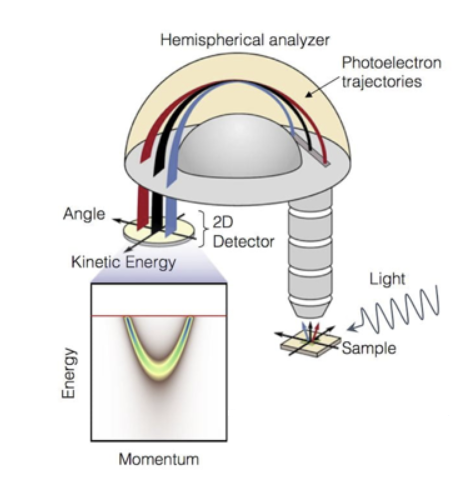}
\caption{\small Schematic of the workings of an electron energy analyzer in an ARPES experiment.}
\label{Holo}
\end{center}
\end{figure}

An illustration of an ARPES experiment is shown in Figure \ref{Holo}. Photoelectrons are emitted from the sample (grey disk) after photon absorption. An electrostatic lens system then sorts these photoelectron into beams with respect to their $\theta$ values, delivering them at a slit that forms the entrance to the energy analysis part of the experiment. 
The hemispherical energy analyzer then disperses the photoelectrons in the direction perpendicular to the entrance slit, such that a 2D image of photoelectron signal vs. $\theta$ and $E_{\mbox{\tiny{kin}}}$ is formed on the detector.
Kinematic and energy conservation considerations enable straightforward conversion of this $\theta$ and $E_{\mbox{\tiny{kin}}}$ information to the number of photoelectrons versus binding energy (or $E-E_F$) and the {\bf k$_{x,y}$} wave vectors relevant for quasi-2D electronic states such as those of the cuprate strange metals investigated, for example, in (Smit et al.~2024).

The next step is to relate these practically measurable quantities to the microscopic properties of the underlying material. We define the current of photoelectrons emitted from the sample as $I({\bf k}, \omega)$, with momentum  {\bf k} and frequency $\omega$, and this depends on three main factors as follows:
\beq
I({\bf k},\omega) = I_0({\bf k},\nu) A({\bf k},\omega) f(\omega)\,.
\eeq
The last term, $f(\omega)$, is the Fermi-Dirac distribution. The first term, $I_0({\bf k},\nu)$, is proportional to the square of the magnitude of the one-electron matrix element that links the initial and final states: $|M_{i,f}|^2$. This involves {\bf k}, the photon energy $\nu$, and is sensitive both to details of the experiment such as the choice of photon energy, intensity and polarization of the light, geometry of the experiment, as well as to microscopic quantities such as the identity of the atomic orbitals from which the band structure is built up.

It is the ability of ARPES to provide access to a measurable signal proportional to the second factor in Eq. (1), $A({\bf k},\omega)$ - known as the single-particle spectral function - that places it centre-stage, as this yields direct information on the dispersion relation and the lifetime of the electronic states of even a strongly interacting electron system such as the strange-metal cuprates. Note that $A({\bf k},\omega) f(\omega)$ is precisely equal to the initial energy-momentum distribution of the electrons to which the probability to emit an electron from the material by a photon should clearly be proportional.

An important property of the spectral function, and one that makes it especially relevant to the semi-holographic modelling of strange metals, is that it can be written in terms of the self-energy $\Sigma({\bf k},\omega)$ of the electron, in the following way:
\beq
A({\bf k},\omega)= -\frac{1}{\pi}\frac{\Sigma''({\bf k},\omega)}{[\omega - \varepsilon_{ {\bf k}}/\hbar - \Sigma'({\bf k},\omega]^2+[\Sigma''({\bf k},\omega]^2}\,,
\eeq 
where $\varepsilon_{\bf k}$ is the initial energy of the electron, $\Sigma'({\bf k},\omega)$ is the real part of the self-energy, and $\Sigma''(\omega,{\bf k})$ is the imaginary part. Hence, $\Sigma({\bf k},\omega) = \Sigma'({\bf k},\omega) +i\Sigma''({\bf k},\omega)$.

By expressing the spectral function in terms of the self-energy $\Sigma({\bf k},\omega),$ we now see how we can connect the holographic modelling of strange metals to experimental data. In particular, by conducting ARPES experiments, we come to know about the spectral function of a metal, and hence about its self-energy. At the same time, we know that we can compute the self-energy, and its real and imaginary parts, from semi-holographic considerations, in particular by relating it to the Green function $G(\omega,{\bf k})$ for the boundary field $\chi$, as we discussed in Section \ref{adscft}. Holographically, this corresponds to a computation in the near-horizon geometry of a charged black hole, as we discussed in Section \ref{semistrange}. In this way, changes in the geometry of the holographic dual to a certain strange metal are mapped to quantities that can be measured experimentally, allowing us to compare the semi-holographic prediction for the self-energy with the ARPES data. 

Recently, Smit et al.~(2024) carried out careful, high-resolution ARPES experiments on simple, single-band cuprate strange-metals, and found that $\Sigma''({\bf k},\omega)$ displays a power law exponent that changes as a function of doping, pointing to a quantum critical phase, and also found that these power law exponents $\alpha(k)$ are also momentum dependent, as predicted by semi-holography, discussed above in Section \ref{adscft}. Thus, semi-holography does indeed provide, at least for the case of the self-energy of the electron, the right description of the appropriate ARPES experiment on a real-life strange metal.


\section{Philosophical issues}\label{philosophy}


Section \ref{semihol} first discusses the relation (and the distinction) between semi-holography and holography, which will be important in the sequel. Section \ref{emergence} then discusses the sense in which the low-energy systems (viz.~the strange metal, and its associated black hole) can be said to {\it emerge}. Here, we view emergence as a balance between (i) novelty and autonomy of one level, relative to another level (usually, a macroscopic level is compared to a microscopic level), and (ii) dependence of the macroscopic level on the microscopic level. This is relevant for semi-holography, because it is a low-energy description that shows novel behaviour and is largely independent of the details of the UV theory, while being rooted in it. Section \ref{explanation} then asks an important question about the relation between the strange metal and the black hole: namely, whether the behaviour of the strange metal is {\it explained} by either the black hole, or by semi-holography. Given the difficulties in modelling and understanding the behaviour of strange metals (see Section \ref{experiment}), it is important to have a clear idea of the degree to which semi-holography does, or does not, explain. Section \ref{realism} then assesses the status of the black hole that is introduced by semi-holographic methods: the question is whether there are conditions under which one can think of this black hole as a real physical system, or whether it is always a mere calculational device. This question is also prompted by the fact that, even though the ARPES experiments favour one side of the semi-holographic relation, the theoretical description is, at least in some cases, symmetric, so that the two systems appear to be theoretically on equal footing. The question of scientific realism then aims to reconcile these two perspectives.

\subsection{Semi-holography vs.~holography}\label{semihol}

In this Section, we discuss the first philosophical issue that is raised by the modelling of strange metals using semi-holography: namely, the distinction between holography and semi-holography. The first question is whether semi-holography is a kind of holography, or whether it is a distinct formal framework. The reason why we focus on this issue first is that it will prove crucial in understanding the various other philosophical issues: namely, emergence, explanation and scientific realism. 

We begin by (i) discussing the relationship between semi-holography and holography in terms of empirical evidence: under what conditions is evidence for semi-holography relevant for holography? We then go on (ii) to discuss why one might want to take holography and semi-holography to be two distinct formal frameworks. Finally, (iii) we discuss arguments in favour of holography and semi-holography being two species of the same genus, namely, duality.

Note that, by ``distinct formal frameworks'', and by referring to holography and semi-holography as a ``species of duality'', we mean the distinction between a duality and a quasi-duality. For our purposes here, a quasi-duality is a relation that does not map all the boundary degrees of freedom into the bulk, but only some of them.\footnote{In general, a quasi-duality is a map that is close to a duality, but falls short of being a duality either by not being bijective, or by not respecting all the structures. Here, being ``close to a duality'' can be either by an approximation or limit (e.g.~tuning a parameter) or by focussing on a subset of states or of quantities. For a discussion, see De Haro and Butterfield (2025). Eskens (2025) develops a notion of effective dualities through a rigorous account of approximate isomorphism and limiting operations.}

Thus our question in this section can be rephrased as follows: whether semi-holography is a duality, like holography, or (as it at first sight appears) a quasi-duality.

A further related point is as follows: recall, from the discussion in Section \ref{theory}, that the bulk and boundary theories related by semi-holography only apply for a specific set of parameters, i.e. in the IR regime of the strange metal, where they give a universal description of its low-energy physics. In other words, they give an effective, low-energy description of the microscopic physics of the strange metal. In the philosophy of duality, this amounts to saying that semi-holography is an effective duality (or, if semi-holography is a quasi-duality, an effective quasi-duality), namely, a relation that approximates a duality in a regime of parameters: in this case, in the IR, and thus falls short of being a duality.

However, note that given the current lack of a clear picture for the microscopic physics of the strange metal, we will only focus on philosophical issues concerning the dual theories related by semi-holography in this paper, and ignore their relation to the more fundamental, microscopic physics of strange metals. Such investigation goes beyond the scope of the present paper (and beyond the state-of-the-art of research in physics), and thus we will set it aside, so as to now discuss our three points, one by one:\\
\\
(i)~~{\bf Semi-holography and empirical evidence.} In distinguishing holography from semi-holography, there is a central point in our exploration of the foundations of semi-holographic modelling of strange metals, 
beyond its intrinsic interest as a novel area of physics research: the possibility of using experimental relevance from strange metals to gain empirical evidence for the physical significance of a holographic theory, in the sense that the experiments could provide evidence for the behaviour of a physical system whose modelling depends crucially (perhaps even essentially) on a holographic theory. In what follows, we will dub this type of evidence {\it empirical evidence for holography}.\footnote{Note that one might worry that there is an ambiguity here: the evidence that we are taking as evidence for holography could just as well be taken as evidence for the field theory side of the holographic duality. From our perspective, this rests crucially on the question about the explanatory role of holography in modelling strange metals: a question that we will return to in Section \ref{explanation}. In the rest of this section, we will work under the assumption that it makes sense to speak of empirical evidence for holography, and postpone discussions of explanation in holographic strange metals until Section \ref{explanation}.} 

Our argument is as follows: the claim that strange metals can provide empirical evidence for holography, in the sense just specified, is based on the use of holography to model strange metals: and so, if laboratory data show key characteristics of the self-energy of strange metals that require holographic methods for their successful description, then it is at least prima facie reasonable to take this as evidence for holography. More precisely, if holography plays a crucial, ineliminable role in deriving and explaining the experimental evidence for strange metals, then we take this to be evidence that, 
at least partially, confirms the holographic framework used to model strange metals: a claim which we take to be in line with standard reasoning for other scientific theories, where if a theory plays a crucial role in explaining a certain piece of empirical evidence, then that evidence is taken to at least in part count in favour of that theory.

However, if semi-holography really is a distinct formal framework from holography, then there is no obvious reason to think that the empirical support for semi-holography, gained from strange metals, should extend to holography. Thus we cannot hope to use strange metals to gain insights into the structure of holographic (as against semi-holographic) theories in particular, and of quantum gravity in particular.\\
\\
(ii)~~{\bf Semi-holography is not holography.} A straightforward argument, though one that we will criticize in point (iii) below,\footnote{Indeed, in later sections, both (ii) and (iii) will be considered and studied with regard to their implications for various philosophical issues.} towards the claim that semi-holography and holography are indeed two distinct formal frameworks, is as follows. As we saw in Section \ref{theory}, semi-holography involves a holographic map that only involves a subset of the boundary degrees of freedom, rather than all of them. For this reason, semi-holography is an asymmetric relation, and cannot be represented as a straightforward equivalence between the bulk and the boundary. On the other hand, holography does involve all of the boundary degrees of freedom. Hence, it is a symmetric relation, and it can be straightforwardly represented as an equivalence relation. Therefore, since a symmetric and an asymmetric relation cannot be the same relation, holography and semi-holography encode different relations between the bulk and the boundary, and so they are two different formal frameworks. In particular, it follows that semi-holography is best seen as a quasi-duality, which, as we have discussed above, does not map all the boundary degrees of freedom into the bulk.

Indeed, to further expand upon the previous argument, one might observe that dualities are understood in the literature as encoding symmetric relations between theories, and in particular they are equivalence relations (De Haro and Butterfield 2025:~Chapter 2). Hence, in conjunction with the above discussion, it follows that while holography is a duality, semi-holography strictly speaking, is not, but, as we said, is rather a quasi-duality; thus confirming our initial suggestion that the two relations are to be kept distinct and cannot be part of the same formal framework.\\
\\
(iii)~~{\bf Semi-holography is holography.} While the observations made in (ii) might seem to endanger the use semi-holography to model strange metals, and thereby gain empirical evidence in favour of holography, it is not clear that the distinction between holography and semi-holography that we just articulated is as stark as it at first sight appears. Thus, in spite of our previous discussion, two main arguments can be made in favour of taking semi-holography to be a kind of holography.

The first is based on an observation in Faulkner and Polchinski (2011), who point out that although the boundary fermion central to the semi-holographic modelling of strange metals does have a holographic dual in principle, this holographic dual is not included in the semi-holographic setup because its gravitational dual would be highly quantum, and so useless for computational purposes. The reason for this lies in the fact that semiclassical gravitational fields are dual to composite operators on the boundary theory; however, the boundary fermion is not a composite operator, and so its bulk dual cannot be a semiclassical field: the bulk dual must be a highly quantum configuration. This argument points to the fact that the asymmetry in semi-holography is of a fundamentally pragmatic character, having to do with what is computationally useful, and less with the fundamental structure of the bulk-to-boundary map. In particular, what this argument indicates is that we can have two theories that are holographic duals, but which look semi-holographic {\it for all computational purposes}, since the degrees of freedom necessary to make the bulk-to-boundary map a duality are highly quantum and therefore not useful for computational purposes.

Another argument for there being a closer relation between semi-holography and holography rests on our discussion in Section \ref{theory}. Recall that there are purely holographic derivations of the semi-holographic setup (G\"ursoy et al. 2012); these constructions allow us to understand the semi-holographic modelling of strange metals in purely holographic terms, thus restoring the duality between the bulk and (a different theory on) the boundary. In particular, such constructions imply, analogously to the argument given above, that the apparent asymmetry between the bulk and the boundary of semi-holography is ultimately a pragmatic matter, stemming from our choice of a specific regime in which to study the relation between the bulk and the boundary, but without any fundamental import on the nature of the relation between the bulk and the boundary theories. 

These two arguments suggest that the initial concern discussed in (i), that the experimental relevance for semi-holography is irrelevant for holography, is not so worrisome after all. For, while the two formalisms appear prima facie distinct, the use of a holographic framework with an alternative choice of boundary conditions brings them together, and we can see semi-holography as a kind of holography. 
In the following sections, we will discuss how other philosophical issues about semi-holography depend on this ability or inability to subsume semi-holography within a holographic framework.

\subsection{Emergence}\label{emergence}

This section focuses on the question of the relation between semi-holography and emergence. The notion of emergence is often used to explain the relation between a theory for a macroscopic system that exhibits complex behaviour, and the theory for its microscopic components. Thus introducing a conception of emergence helps us to understand the relation between these two levels as a balance between: (i) the novelty and autonomy of the macroscopic level with respect to the microscopic level; and (ii) the dependence of the macroscopic level on the microscopic level.\footnote{For recent discussions of emergence in the philosophy of physics literature, see for example Butterfield (2011a,b), Guay and Sartenaer (2016), Humphreys (2016), Crowther (2016) and De Haro (2019).}

In the case of semi-holography, the main distinction between levels will be between the high-energy UV theory and the low-energy IR theory.\footnote{Additionally, in Figure \ref{Emergencefig2} below, we also discuss horizontal emergence, between two low-energy theories.}
The emergent description is then rooted in a microscopic UV theory, some of whose {\it general} features are known: but the IR theory is independent of the details of the UV theory, and it is robust with respect to small changes in the UV. This then illustrates the motivation we gave in the preamble of this section: the notion of emergence allows us to understand the rootedness of the macroscopic level in the microscopic level, without having to worry about all the microscopic details.

We begin by recalling a standard conception of emergence. Then we will use this conception to identify three different cases as cases of emergence which are depicted in turn in Figures \ref{Emergencefig1}, \ref{Emergencefig2} and \ref{Emergencefig3} below.\\
\\
{\bf A conception of emergence and an example.} We kick-off by using the conception of emergence by Butterfield (2011a:~p.~921), which has been very influential in the recent philosophy of physics literature:
\begin{quote}\small
I shall take emergence to mean: properties or behaviour of a system which are {\it novel} and {\it robust} relative to some appropriate comparison class. Here ``novel'' means something like: ``not definable from the comparison class'', and maybe ``showing features (maybe striking ones) absent from the comparison class''. And ``robust'' means something like: ``the same for various choices of, or assumptions about, the comparison class'' ... I shall also put the idea in terms of theories, rather than systems: a theory describes properties or behaviour which are novel and robust relative to what is described by some other theory with which it is appropriate to compare.
\end{quote}
Thus in Butterfield's account, the distinctive features of emergence are {\it novelty} and {\it robustness}, relative to an appropriate comparison class. Also, emergence applies to both systems and theories (including the {\it properties} that theories describe): and we will adopt both usages of the word ``emergence'' below. To illustrate this conception of emergence, we use a well-known example:\\
\\
{\it The example of water: novelty and robustness.} A well-known example of emergence is the formation of liquid water from the underlying water molecules, when the spacing between them is very small, much smaller than the wavelength of the excitations of liquid water.\footnote{In addition, other factors such as inter-molecular bonding, which is weak individually yet collectively significant, must also be considered.} 
Rephrased in terms of theories: hydrodynamics can be seen to emerge from the underlying molecular theory by - under certain conditions - taking an appropriate limit. To unpack this claim, we need to discuss the novelty and robustness of the emergent properties: this will also help us to illustrate the relevance of the notion emergence for semi-holography later on.

Hydrodynamics has {\it novel properties} relative to the comparison class (viz.~the molecular theory), for example in that it is a continuum theory that describes water as having macroscopic properties such as viscosity. These properties are novel because they are absent from the comparison class (namely, the molecular theory is not a continuum theory, surely not in the sense of hydrodynamics; nor does it have a viscosity). 

The properties of water are also {\it robust}, because they are the same for various choices of the comparison class. For example, there are many configurations of the water molecules that give rise to the same hydrodynamic state. Thus, since emergence relates many microscopic states (i.e.~a whole comparison class) to a single macroscopic state with emergent properties, emergence is an {\it asymmetric} relation. More generally, we cannot exchange the theory that emerges and the theory (or set of theories) that it emerges from, and that form the comparison class.\\
\\
{\it Emergence and reduction are compatible.} In principle, one can of course derive the continuum properties of water from the microscopic theory by taking a limit, and calculate the value of the viscosity in terms of the parameters of the microscopic theory, such as the density of the molecules, the temperature, etc.. Yet the hydrodynamic behaviour is novel: and, in this sense, Butterfield (2011a,b) and others have cogently argued that emergence and reduction are compatible: novelty of behaviour is the main feature of emergence, and it is compatible with the mathematical derivability of one theory from the other.\footnote{This notion of emergence has been further developed in De Haro (2019), where novelty is explicated as novelty of reference, which can be cashed out in terms of a criterion, that the linkage and interpretation maps do not commute.} 

Having introduced emergence in general, we now turn to semi-holography:\\ 
\\
{\bf Emergence and semi-holography.} There are three main questions about emergence in connection with semi-holography in the context of strange metals:

(A)~~whether there can be emergence of the bulk theory from the boundary theory, when these are also duals (i.e.~in our case (iii) from Section \ref{semihol}); 

(B)~~whether the strange metal emerges in the IR; and 

(C)~~whether the black hole emerges.\\
\\
Question (A) is a general point about the relation between emergence and duality, which we will argue are at odds with one another. This implies that, when answering questions (B) and (C), we need to first resolve this tension. To do that, we will build on the distinction made in Section \ref{semihol}, between the opposing cases that semi-holography \textit{is} holography, and that it is \textit{not} holography. Question (C) is in part motivated by the general idea of spacetime emergence in quantum gravity, as in discussions of gauge-gravity dualities it is sometimes said that the bulk spacetime emerges from the CFT. The question then is whether something similar can (or cannot) be said about semi-holography.

The above list of questions and the presented cases of possible emergence are not meant to be exhaustive. Also, our aim here is not to give a detailed answer to each of them: rather, as we discussed in Section \ref{intro}, we list the most significant and interesting cases of emergence and, by using tools that have been introduced in the recent literature on dualities and emergence, we sketch answers to the above questions in the context of semi-holography. 

Consequently, in answering these questions, we will focus on three emergence scenarios that we label as: \textit{vertical}, \textit{horizontal} and \textit{diagonal}, depicted below in Figures \ref{Emergencefig1}-\ref{Emergencefig3}, respectively. The vertical and diagonal cases, both possess a vertical component, assume that semi-holography is a duality because it can be successfully derived from a holographic theory (see point (iii) in Section \ref{semihol}). In the horizontal emergence scenario, semi-holography is not a duality, i.e.~that the reduction is not successful, which we recall - from (ii) in Section \ref{semihol} - is an open question.
\\
\\
So, we now go on to answer the three questions (A)-(C) posed above:\\
(A)~~{\it Emergence is at odds with duality: no horizontal emergence.}

First, we note that there is an objection to having emergence across a duality relation, as a duality is a symmetric relation (see (ii) in Section \ref{semihol}), while emergence is an asymmetric relation, as described in the last paragraph of the discussion of the example of water.
Therefore, a single relation cannot be a case of both duality \textit{and} emergence\footnote{For a detailed discussion of this argument, see Dieks et al.~(2015) and De Haro (2017).}
and if semi-holography is a duality, then the conclusion would be that there is no horizontal emergence, i.e.~no emergence ``in the direction of the duality relation''.

If one wanted to deny this conclusion, then one first either needs to deny the premise that semi-holography is holography, or secondly one would need to deny the horizontal nature of the emergence itself. This amounts to breaking with the general assumption that emergence and duality have to be the same relation. 
In the former, semi-holography is then only a quasi-duality, and since this is an asymmetric relation, it is compatible with emergence, and therefore we can combine emergence with semi-holography. We will return to this case again in our discussion of the answer to question (C), below.
The latter is in line with the holographic derivations of semi-holo\-graphy discussed in point (iii) of Section \ref{semihol}, which - by making semi-holography a duality - brings semi-holography back into alignment with holography. This does mean that we need to find a way to distinguish semi-holography from emergence, and consequently we require a different relation (of emergence) to the horizontal case, and a ``vertical'' component needs to be added.
\\
\begin{figure}
\begin{center}
\includegraphics[height=3cm]{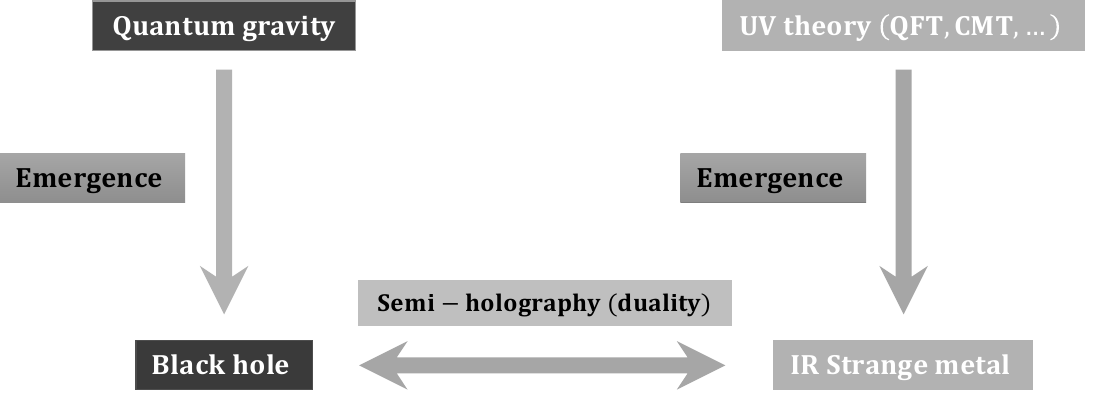}
\caption{\small Emergence beside semi-holography: they lie along different axes, as in the discussion of question (B).}
\label{Emergencefig1}
\end{center}
\end{figure}
\\
\\

(B)~~{\it Emergence beside semi-holography: vertical emergence.}

To address this question, we discuss a second way in which a duality can be combined with emergence: namely, there can be emergence if - in addition to the two theories being duals - each dual has a parameter such as a distance, or an energy scale that can be used to ``tune'' the emergent behaviour.
This means that emergence takes place along the direction of the tuning parameter and since duality and emergence are now ``along different directions'' as illustrated in Figure \ref{Emergencefig1}, they are not at odds with one another.\footnote{De Haro and Butterfield (2025:~\S 14.3.4) call this option `emergence beside duality'.}
In holographic dualities, the energy scale at which we probe the CFT, namely in the UV or in the IR regime, would be a natural candidate for such a tuning parameter. 

Consequently, there is a straightforward sense in which, similar to the emergence of the macroscopic properties of liquid water from the underlying molecular dynamics, we can speak of the emergence of strange metal properties (such as the linear dependence of the resistivity on the temperature and the momentum dependence of the self-energy) in the IR.
These emergent properties are novel, because they are in general not possessed by the UV theory.
They are also robust, as they are independent of the details of the UV theory because there can be many different quantum field theories or condensed-matter theories in the UV that lead to the same IR behaviour.
\footnote{Since question (B) is about the emergence of the strange metal, the consideration of Figure \ref{Emergencefig3}, which also contains a vertical direction along which the strange metal could emerge, does not add anything new here. The reason that Figure \ref{Emergencefig3} distinguishes itself from Figure \ref{Emergencefig1}, is in the emergence of the black hole, which is relevant for the discussion of question (C).}\\
\begin{figure}
\begin{center}
\includegraphics[height=1.5cm]{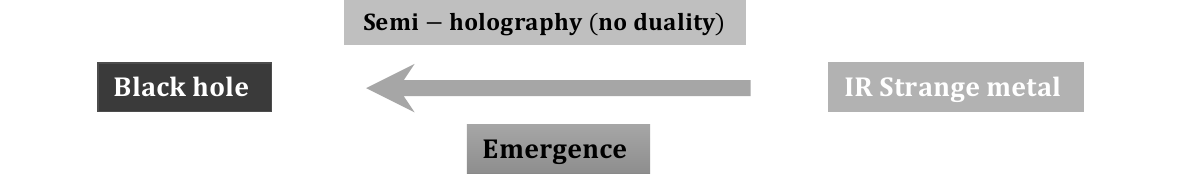}
\caption{\small Emergence of the black hole from the strange metal, under the assumption that semi-holography is not holography, as discussed in the context of question (C). In this case semi-holography is not a duality.}
\label{Emergencefig2}
\end{center}
\end{figure}
\\
(C)~~{\it Horizontal and diagonal emergence.} Unlike the emergence of the strange metal, where our answer to question (B) only considered vertical emergence, there are in principle two routes by which the black hole could emerge from the QFT: either in the horizontal or in the diagonal direction. The distinction between these two routes follows from distinguishing between the cases that semi-holography is not, or is holography. 

If semi-holography is not holography (in the sense of point (ii) in Section \ref{semihol}), then we can have emergence of the black hole in a straightforward sense: namely, the black hole emerges as a low-energy description of the strange metal, similarly to how liquid water emerges from the underlying molecular interactions and structure. In this situation, regardless of the specific UV completion, the black hole gives an effective and very useful description of the semi-holographic system, illustrated in Figure \ref{Emergencefig2}. 

However, if semi-holography were to be holography, then again we would need to deal with the tension between emergence and duality as discussed above in the context of question (A).
Since there is a duality in the IR, so that the black hole is - within a certain regime of parameters - dual to the strange metal, then neither the black hole nor the strange metal can emerge from the other. They are equivalent descriptions of the same ``invariant physics''.
This ``invariant physics'' is described mathematically by the set of correlation functions of the theory in the IR, and what the gravity dual adds to this is a connection to a classical gravity theory that enables the interpretation in terms of the horizon of a black hole.\footnote{Several authors have stressed the idea of a `common core theory' behind two duals: see, for example, De Haro and Butterfield (2025:~Sections 1.1, 2,2, 12.2, 12.4) and Le Bihan and Read (2018).}

\begin{figure}
\begin{center}
\includegraphics[height=3cm]{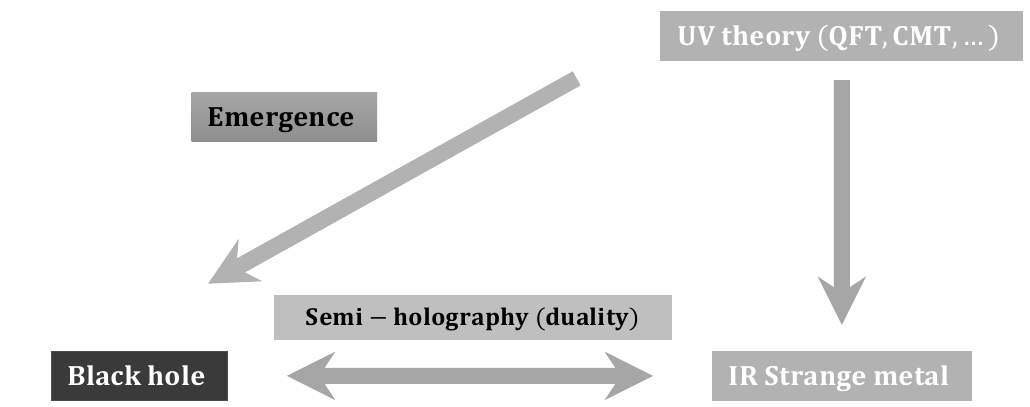}
\caption{\small Emergence of the black hole from the lower-dimensional theory in the context of question (C), thanks to a combination of the RG flow, and semi-holography. Here, semi-holography is assumed to be holography.}
\label{Emergencefig3}
\end{center}
\end{figure}

This approach - in order to get interesting emergence of the black hole across the duality - necessitates consideration of the energy scale of the dual CFT, as we also did in the discussion of question (B).
As Figure \ref{Emergencefig3} illustrates, the black-hole horizon can then emerge from the UV QFT by a combination of firstly the renormalization group (RG) flow towards the IR, and, secondly the duality map that allows the reinterpretation of the strange metal as a black hole.
The properties of the black hole, such as the presence of a horizon, are clearly novel with respect to the UV QFT, and they are also robust, in that they are independent of the UV completion for the same reason as before: namely, that the emergent IR physics is independent of the UV. 

Note that the difference between Figures \ref{Emergencefig1} and \ref{Emergencefig3} is in what Butterfield's conception of emergence calls the {\it comparison class}. In Figure \ref{Emergencefig1}, the comparison class is a gravitational theory (some theory of quantum gravity, such as string theory, for example). In Figure \ref{Emergencefig3}, the comparison class is a lower-dimensional theory in the UV. This can be a conformal field theory, a condensed matter theory, or some other quantum theory: we can have very different theories here, as the IR behaviour is largely independent of the details of the UV.

In Section \ref{realism}, we will return to the question of the realist interpretation of the emergent physics.

\subsection{Explanation}\label{explanation}

The last two sections pointed to interesting puzzles in the foundations of semi-holography, both of which have \textit{explanation} in common. 
Point (i) in Section \ref{semihol} can be seen as being about whether or not holography explains the empirical evidence for semi-holography, when viewed through the lens of whether experimental evidence for semi-holography is relevant to holography.
In a related manner, Section \ref{emergence}'s focus on emergence, naturally ties into questions of explanation, since the use of the notion of ``emergence'' itself arises, as we have stressed, from a desire to explain.
On the one hand, emergent phenomena are taken to have a degree of novelty and explanatory autonomy from their fundamental constituents. On the other hand, establishing that one theory emerges from another theory is usually understood as providing an explanation for the features of the emergent theory that we did not have before, i.e.~one that is given in terms of its microscopic constituents. 

Therefore, this section focuses on explanation in the context of the semi-holographic modelling of strange metals.\footnote{For an introduction to philosophical issues regarding explanation, see either Woodward and Ross (2021), or for a more in-depth discussion, Chapter 6 of Curd and Cover (1998).} 
Our goal here is to understand, under which conditions holography can be said to play an explanatory role for the behaviours of the strange metal that are well-described by holography, rather than it being merely an efficient computational device. In the following, we will explain (a) why one might be sceptical about the explanatory value of holography; (b) how one might overcome such scepticism and judge holography to be explanatory of relevant strange-metal data; and (c) how this task crucially depends on whether semi-holography is a kind of holography, which formed the core of Section \ref{semihol}.

Note that we focus in this section on the possibility of using holography to explain the properties of strange metals, rather than focusing on the opposite explanatory direction from holography (and field theory/strange metals) to the gravitational black-hole dual. \footnote{The opposite direction of explanation is more natural in the context of quantum gravity, where one is interested in understanding gravitational phenomena through their field theory duals. However, the gravitational dual to a strange metal is not a realistic candidate for describing the physics of our world, since it is a five dimensional black hole in a spacetime with a negative cosmological constant; in particular, questions of how holography can explain empirical evidence, as we are considering in this Section, are a no-go for such a system, since there is no possible empirical evidence available for it.} 
Explanations of empirical evidence from holography and field theory to gravity would be more natural and interesting if one were to find candidate gravitational duals that might be realized in our world; however, this is not currently the case for field theories describing strange metals.\footnote{Note that while this is not the case for strange metals, the development of holographic models for cosmology is an active area of research, which has seen significant research development. For a sample of the state of the art in the field, see, e.g. Strominger (2001), Witten (2001), Chakraborty et al. (2023), Chadrasekaran et al. (2023).}
Nevertheless, the following Section \ref{realism} will provide a brief discussion of this ``to the black hole dual'' option.
\\
We make this choice for the ``to the strange metal dual'' as the direction of explanation in this section as this is the natural one insofar as one is interested in how holography can relate to the concrete experimental results described in Section \ref{experiment}, and more generally how holography enters into the modelling of strange metals. \\

(a)~~\textbf{Against explanation in semi-holography.} We begin by recalling why one might be sceptical about the possibility 
of treating the gravitational dual of a strange metal as being 
explanatory, in the usual substantive sense of the word. The reason for this is intuitively clear: on major accounts, to speak of an explanatory relation, the explanation itself is required to be true.\footnote{If one's account of explanation does not require the factivity of the explanans and-or the explanandum, then this sceptical argument of course does not arise.}
In the case of a strange metal in a laboratory here on Earth, this would mean that were the gravitational dual to be explanatory, it must be true that a five-dimensional black hole is involved in the ARPES experiments on strange metals described in Section \ref{experiment}. However, it seems highly implausible that the experimental apparatus described in Section \ref{experiment} can actually accommodate the existence of a five-dimensional black hole, as would be required to make the gravitational dual explanatory (see also the discussion of point (i) in Section \ref{realism}). Hence, the higher-dimensional black hole dual to the strange metal in our semi-holographic setup cannot play an explanatory role in the experiments described in Section \ref{experiment}.

The main issue with this conclusion is that it reduces the whole project of modelling strange metals using holography to an exercise in identifying a more efficient method to compute observables, with no explanatory power. 
While this might indeed be the case, it is both at odds with a realist perspective on scientific theories (more on this in Section \ref{realism}), and would cast holography in the form of a disappointingly ``black box'' that leaves us in the dark as to why strange metals actually display such behaviour.
Consequently, in the following, we take some effort to try and identify under what conditions we might avoid this conclusion and be able to regard the construction of holographic models of strange metals as explanatory.\\
\\
(b)~~\textbf{How holography can explain properties of strange metals.} 
The solution to the thorny question outlined above actually appears to be straightforward: we need a description of the semi-holographic setup that retains the features of holography, without committing ourselves to the truth of implausible claims about the gravitational dual being involved in the strange-metal experiments themselves.

The natural candidate for such a description is one that is well-known from the philosophy of dualities in which a common core theory underlies the two duals (De Haro and Butterfield 2025). This theory is defined only in terms of quantities invariant under the duality map, in this case the bulk-to-boundary map.
The reason why we take the common core theory to solve our predicament and guarantee the explanatory value of holography in modelling strange metals is due to the fact that it deals only with invariant elements, eschewing structure like the black hole itself (which is clearly not invariant between the bulk and the boundary), while preserving the basic structure of holography. It can do this since it does encode the fundamental features of holographic theories, namely those features that do not depend on the specific structure of the dual theories. 

Thus insofar as we can describe the semi-holographic modelling of strange metals in terms of the common core theory, we have a natural candidate for an explanation of the behaviour of strange metals that does not commit us to the truth of implausible claims: indeed, to the truth of any claim that is not already made true by the boundary theory itself (De Haro 2023) and hence acceptable for purposes of explaining strange metals.\\
\\
(c)~~\textbf{A common core requires a duality.} The previous point appears to offer some hope for holographic approaches to be more than computationally efficient black boxes. However, an important point to consider is under which conditions a common core theory would be available to us. 
Here we need to recall that in semi-holography, we have an asymmetry between the bulk and the boundary: the boundary fermion lacks a bulk dual, suggesting that a common core theory is not available.
Consequently, if we tried to define a common core theory in the semi-holographic context, the boundary fermion would not be included in it, since it does not have a bulk dual and it itself is not preserved across the duality. In this way, we lose crucial information necessary to model strange metals, as illustrated by that fact that no arrow points to the ``IR strange metal'' in Figure \ref{Emergencefig2}).

If we want to have a common core theory, we need to look at the holographic derivations of semi-holography discussed in Sections \ref{theory} and \ref{semihol}, as suggested by Figure \ref{Emergencefig1}.
In particular, approaches to semi-holography that restore a duality between the bulk and the boundary, would allow the definition of a common core theory which does not lose any information essential to modelling strange metals.\footnote{At least, such approaches would allow this conditional on the constraints on the existence of a common core familiar from the philosophy of duality, see, e.g. De Haro and Butterfield (2025).}
In the next Section, we will discuss these kinds of common core theories and their interpretation in more detail.

\subsection{Scientific realism}\label{realism}

In Section \ref{emergence}, we discussed the emergence of a black hole in the IR limit of a theory of quantum gravity (Figure \ref{Emergencefig1}) and of a quantum field theory of a strange metal (Figure \ref{Emergencefig2}). This was followed in Section \ref{explanation} by the question whether this black hole can have an explanatory role or whether we should see it only as an efficient calculational device. 
Both the \textit{emergence} and \textit{explanation} discussions prompt a question about the interpretation of this black hole: are there conditions under which we can take the existence of this black hole seriously, or is it a convenient instrument for calculation? 
We can add a third motivation for this question from Section \ref{theory}, based on the case that semi-holography \textit{can} be reduced to holography, resulting in a symmetric relation, despite the obviously enormous physical and experimental/observational differences that we expect between the two duals of a little piece of strange metal one can hold between finger and thumb and a higher-dimensional black hole.
Thus, we need to confront the question which of these two systems (or both) is (are) described by a semi-holographic duality? Or, put differently:
\begin{itemize}
    \item Given the evidence of the ARPES experiments, can a realist interpretation of the black hole be upheld?
\end{itemize}
By ``realist interpretation'', we mean an interpretation that (under the conditions for which the theory says that there is a black hole) warrants the belief in the existence of this black hole.\footnote{As in discussions of scientific realism, by having ``warranted`' belief, we mean having epistemic justification for it.

As we forego discussion of the details of or specific types of scientific realism here, we refer the reader to Chakravartty (2017) for a general introduction, and to e.g.~Matsubara (2013), Le Bihan and Read (2018), Rickles (2011), and De Haro and Butterfield (2025:~Chapter 13) for discussions in the context of dualities.}

In this section, we discuss three objections against such a realist interpretation, to which we will then also furnish replies which emphasize that, even though a realist view can perhaps be given consistently, in its present form it does entail an arbitrary interpretative choice, undermining the degree to which this view is convincing.

In philosophy, the questions of explanation and scientific realism are usually closely connected, so the arguments here mirror and further develop those about explanation given in Section \ref{explanation}. There, we concluded that the black hole is itself not explanatory, although the common core can have an explanatory role, with semi-holography being the thing that is to be explained (known in philosophy as the explanandum).
In the following, we first present a trio arguments as to why there is no straightforward sense in which we can say that the black hole described by semi-holographic theories exists. Then, in the replies to these objections, we argue that the only viable realism is not realism about the black hole itself, but about the entities described by the common core theory.\\
It is fairly clear that the general answer to the question bulleted above is \textbf{No}. In cases of semi-holography, one is not justified in asserting that there is a black hole. 
There are three main reasons for this negative answer, i.e.~three objections to a realist interpretation of the black hole, each of them relevant in all three cases of emergence illustrated in Figures \ref{Emergencefig1}-\ref{Emergencefig3}.\\
{\bf Three objections and replies:}\\
(a)~~The black-hole solutions of semi-holographic theories are embedded in anti-de Sitter spacetime (and sometimes they are also higher-dimensional), while the universe around us is not an anti-de Sitter spacetime. 

(b)~~Strange metals \textit{are} realized in the lab, and yet there is clearly no black hole there. Thus a realist interpretation is inconsistent with the empirical evidence.

(c)~~If the quantum field theory on the boundary is one's fundamental theory, then in general the black hole in anti-de Sitter spacetime is not supposed to give a realistic description of a strange metal: it is clearly an effective instrument used to enable calculations that are otherwise very difficult, or even impossible to do with our current knowledge of strongly and long-range  entangled fermions. So, if the aim is to describe a strange metal, the gravity theory was never meant to be a realistically interpreted theory in the first place. 

This third argument gets additional support from the fact that, if semi-holography is not a duality but only a quasi-duality, as in Figure \ref{Emergencefig2}, then it is not just that the black hole was not {\it meant} to give a realistic description of the physics, but also that it {\it cannot} give a realistic description, since it misses degrees of freedom that {\it are} described by the boundary theory.

These objections distinguish different aspects of our sceptical remark (a), in Section \ref{explanation}, about the black hole being explanatory. The first two objections just presented [(a) and (b)] appeal, respectively, to the cosmological and the experimental evidence (cf.~Section \ref{experiment}). Objection (c) is about the instrumental nature of the black-hole system, and its inability to describe the system in the lab. These three objections are particularly strong for the question of a realist interpretation of the black hole. 

However, if we were to close our discussion here, at this inability of justifying, at this very general level, a realist interpretation of the black hole, then our conclusion would not only be dull: it would also not do justice to the role that these kinds of discussions play in physics, where they serve as a tool helping construction of viable interpretations of our experiments and theories in a variety of situations.
Thus, we take on the footwork to investigate whether there can be cogent replies formulated to each of the objections just raised.

Broadly speaking, we will find no credible realist defence of the black hole's existence: and the different reasons for this failure of a realist defence will be instructive. 

In the case of horizontal emergence, i.e.~Figure \ref{Emergencefig2}, the reasons are straightforward: here, realism about the black hole is compatible with neither the experimental, nor the theoretical, evidence. Thus, in this case, while there can  be a reply to objection (a), no reply to objections (b) and (c) can be formulated.

When emergence has a vertical component, as in Figures \ref{Emergencefig1} and \ref{Emergencefig3}, the reasons for rejecting a realist view of the black hole are of a more subtle epistemic nature, relying on our requirement that scientific realism be {\it cautious}. Cautious scientific realism is not a specific kind of scientific realism, but rather a set of epistemic and logico-semantic arguments that, in the face of dualities and other inter-theoretic relations, recommend a cautious attitude towards the interpretation of scientific theories.\footnote{See M\o ller-Nielsen and Read (2017) and the development of the cautious realist view in De Haro and Butterfield (2025:~Chapter 13).}

As we will argue, a cautious realism recommends the {\it rejection of instrumentalism} as regards either side of a duality. We will argue that, although a `black-hole monomaniac' can by pure logic uphold their belief in the existence of a black hole, the flipside of this attitude is an unwarranted instrumentalism as regards the dual.\footnote{In the context of quantum mechanics, Butterfield (2021:~p.~66) has called the interpreter who insists on advocating position, at the expense of momentum, a `position monomaniac' and the advocate of the dual, i.e.~of momentum, is a `momentum monomaniac'. In our discussion here, we set aside other details, such as the possible embedding of the black-hole solution in higher-dimensional string theory or M-theory compactified to five or four dimensions, since it does not change our discussion of the compactified solution.\label{monom}}  
Thus, absent additional argumentation, the monomaniac is not justified in their belief: while, perhaps surprisingly, the caution that is the characteristic of Buridan's ass is commendable here, for a different reason than usually considered. (The reason why this caution is commendable, absent in the usual discussion of Buridan's ass, is that it has a better alternative to a choice between duals: namely, a common core theory.)\\
\\
After this preamble, we now turn to the replies to the objections:\\
\\
{\it Reply to objection (a)}. This is a fairly standard issue in discussions of scientific realism:\footnote{For one such view, where the realist's interpretative task is to articulate the possible world that is described by the theory, even if this does not always coincide with the actual world, see e.g.~North (2021:~p.~187).}
a realist interpretation of a theory (and/or of a specific solution of a theory) is not always meant to give a description of the actual world. Both general relativity and quantum field theories admit many solutions (the G\"odel universe, wormholes, magnetic monopoles, instantons, etc.) that may not exist in our actual universe, but for which natural questions of realist interpretation do arise: is there a possible world as described by a given solution of this theory? What does that possible world look like? This is an interpretative question that we can also reformulate thus: setting aside the empirical evidence for the number of dimensions of our actual world, what is an admissible (or perhaps even a best) interpretation of the given theory? Thus, in all of the cases in Figures \ref{Emergencefig1}-\ref{Emergencefig3}, objection (a) is defeasible (for the physics readers: defined as open in principle to revision, valid objection, forfeiture, or annulment).\\
\\
{\it Reply to objection (b)}. This objection can be circumvented in a way that is suggested by the reply to the first objection: namely, if the strange metal is not part of the actual world, but only part of an idealized ``quasi-metal world'', then (in cases that involve a symmetric horizontal relation, i.e.~Figures \ref{Emergencefig1} and \ref{Emergencefig3}) it is not clear that we can set up any experiment that can tell the difference between a quasi-metal and a black hole.
We cannot bring in and use a system ``from the outside'', not itself described by the holographic theory, to tell the difference between a quasi-metal and a black hole.\footnote{This condition is, in effect, that a common core theory exists that is unextendable: our possible world is all and only the world described by the holographic quasi-metal theory, and we cannot break the duality by extending it.}
In other words, the fact that these systems are realized in the lab, where there is no black hole to be found, does not imply that in a possible world that is described by just the holographic theory, experiments {\it would} be able to tell the difference. For the experiments would themselves be mapped from one dual into the other, just as in Poincar\'e's thought experiment of the disk world.

Note that this reply is not available if, as in Figure \ref{Emergencefig2}, the horizontal emergence relation is asymmetric, and there is no expectation that the possible worlds described by the two theories are the same. Thus in this case there is a clear preference for the strange metal, and no realism about the black hole. (This is then similar to our negative conclusion about explanation (a) in Section \ref{explanation}.)\\
\\
{\it No refutation of objection (c).} This objection, especially its second part about the asymmetry of semi-holography, is substantive, and there is no straightforward reply to it. 

The second part of this objection says that if semi-holography is asymmetric as in Figure \ref{Emergencefig2}, the black hole cannot give a realistic description of the physics, because it does not have enough degrees of freedom: at best there can be some kind of effective realism.\footnote{See Williams (2019).} However, objection (b) already said that there can be no realism about the black hole in the case where Figure \ref{Emergencefig2} is relevant, so that, in this case, {\it instrumentalism about the black hole} is justified. 

This leaves us with the symmetric cases in Figures \ref{Emergencefig1} and \ref{Emergencefig3} as the interesting candidates for realism, also recalling that in point (c) from Section \ref{explanation}, these were also the relevant cases for semi-holography to be explanatory.
In these cases, we can extend the usual discussions of scientific realism for dualities, and considerations about their interpretation, to semi-holography. The justification of this assumption then picks up our theme, from point (iii) in Section \ref{semihol}, that semi-holography is a type of holography and that the asymmetry of semi-holography is a pragmatic matter anchored in the radically different computational difficulties of the two duals for a given regime of parameters. In this case, we can then bring the interpretative resources of the philosophy of dualities to bear on semi-holography.\\
\\
{\bf Using internal and external interpretations.}\\
After having just brought forth the objections to realism and their possible replies, we present one final aspect of the problem before we conclude.
Standard discussions of dualities distinguish between internal and external interpretations: in an internal interpretation, only the common structure of dual theories, and not the specific structure of each of the duals, has physical significance. For example, in so far as the geometric properties of the AdS spacetime do not have a counterpart in the CFT, we do not interpret the theory as describing a curved spacetime of general relativity, i.e.~we do not ascribe physical significance to the AdS spacetime. Likewise, in so far as the local (as against global) gauge symmetries of the CFT do not have counterparts in the bulk, and all of the theory's observables are independent of these local gauge symmetries. We do not take them to be part of what the theory describes: we do not take them to have physical significance.

An \textit{internal} interpretation of semi-holography does not appear to allow a realist interpretation of the black hole. This is because the higher-dimensional geometric structure is not part of what is interpreted by the internal interpretation: at least, not in the straightforward geometric terms familiar from general relativity. Rather, an internal interpretation would seem to privilege a lower-dimensional common core theory, perhaps defined by the set of all correlation functions. This privileged role of the common core echoes our discussion of explanation in part (c) of Section \ref{explanation}.

Note that it is not required that such a formulation be a CFT formulation. Thus, we are not favouring one of the two sides of the duality.
Rather, our current (and, admittedly, defeasible\footnote{We say that this understanding is ``defeasible'', because it depends on what a non-perturbative formulation of (semi-)holographic dualities ends up being, and we do not currently have such a non-perturbative formulation. Thus our point is that the current evidence points towards a formulation of the kind discussed here. Namely, the semi-classical limit suggests that this quantum theory is defined on a relativistic spacetime that is isomorphic to the spacetime of the asymptotic boundary, and so that it is a quantum field theory.}) understanding of the common core theory for AdS-CFT is that it is best formulated as a quantum theory with a Hilbert space defined by the set of all (boundary) correlation functions and that has a notion of weak and strong coupling. Even if some of the geometric structure might be common to the two duals, the internal interpretation seems to be in terms of the semi-classical limit of a strongly coupled quantum theory, rather than in terms of a classical solution of a gravity theory in one extra dimension.

By contrast, \textit{external} interpretations do assign physical significance to the specific structure of duals. These interpretations are therefore more plausible candidates for a realist view of the black hole. For, if one aims to interpret the semi-holographic theory in terms of a possible world, rather than in terms of the actual world (so that one can set aside any measurements of the quasi-metal properties in experiments that are not described by either side of the holographic theory), then one could in principle insist on privileging the higher-dimensional, gravitational, interpretation.

But what would warrant such principled realist commitment to one side of the duality, with the corresponding instrumentalist non-commitment to the other side? We are not aware, in any of the examples of dualities, of a cogent argument (physical or otherwise) that would warrant a commitment to one side of the duality, with all of its specific structure, at the expense of the other side, with its specific structure. 

This need not - of course - prevent a ``black-hole monomaniac'' from insisting on adopting such an external interpretation, and thereby perhaps being able to develop a consistent theory.\footnote{For our use of the phrase ``monomaniac'', see footnote \ref{monom}.}
But a theory can of course be logically and metaphysically consistent while lacking in epistemic warrant (defined as requiring that a belief is formed or held in a way that is conducive to that type of belief being true). In our example, the monomaniac fails to take heed of the duality and its logico-semantic implications: 
the external interpretation depends on an arbitrary choice, 
and thus (at this very general level, absent further argumentation) it is left unjustified.\footnote{Although, as we have stressed, our point here is {\it not} that the inadequacy of an external interpretation of semi-holography follows as a matter of logic, since we cannot exclude that some non-arbitrary reasoning might exist that privileges one dual at the expense of the other dual: we are sceptical that, under the condition that we have stated (namely, having an unextendable common core theory), such an argument can be cogently made. To the best of our knowledge, such an argument has not been made for any of the examples of dualities that satisfy this condition. Thus we judge it to be more cautious (and, by analogy, more in line with the motivationalist approach to dualities that we endorse) to be sceptical about the monomaniac's claim, before the reasons for such a choice are spelled out. This also because, in view of epistemic uncertainty, the advocate of the common core does have principled reasons to privilege an internal interpretation. For a more detailed argument for why a cautious realism warrants a commitment to the common core theory, but not to one dual at the expense of the other, see De Haro and Butterfield (2025:~Section 13.2).}

One question that we have not considered in any detail is how high one can go in the RG flow and still have a duality, as it is generally expected that the duality will break at high energies, even if only because the atomic structure of the metal becomes relevant. Likewise, one can ask how high one can go in temperature---one does not expect the holographic description to be valid near the melting point of the metal. At such points, any condensed matter description will break down. While these are relevant questions for the physics of this system, they are not relevant for the low-energy effective theories that we have considered in this paper on both sides of the duality. In other words, it is reasonable to expect that the holographic map between the strange metal and the black hole is only an effective duality. However, this need not stand in the way of giving explanations using concepts and properties from effective field theory, and having a realist commitment to the entities involved in such explanations. 

To conclude this section: although, in this case, a realist commitment to the common core between duals is clearly justified (and it will result in something like a commitment to a lower-dimensional quantum theory, as defined for example by an infinite set of correlation functions), we argue that, at least based on our present understanding of the common core, there are good reasons for scepticism about a justified realist commitment to the existence of a black hole exactly as envisaged by general relativity. An internal interpretation does not seem to admit an interpretation of the solution as a black hole of general relativity, while one-sided external interpretations do not seem to rest on sound and non-arbitrary interpretative principles, and surely none that we could identify in this paper. If such reasons could be given, then the black-hole realist could have their cake and eat it.

\section{Summary and Conclusions}\label{conclusion}

In this paper we have proposed strange metals to be an interesting arena in which to carry out natural philosophy, in which theoretical and philosophical ideas about holography can make contact with real-world experiments.
In this endeavour, the link between semi-holography and strange metals, and in particular with angle-resolved photoemission (ARPES) measurements of the self-energy of the electrons in a set of differently-doped strange-metal cuprates is both the launchpad and testbed of our natural philosophical discussion.

First-up, in Section 1, we set out the idea behind bringing the physics and philosophical aspects of this problem together, and we set the scene as regards what strange metals are.

In Section 2, we first touch on the canonical (Fermi liquid) theory for electrons in metals, before discussing why a semi-holographic approach is required to describe the ARPES experiments. Then, the relationship between semi-holography and the anti-de Sitter conformal field theory (AdS-CFT) correspondence is discussed in Section \ref{adscft}, prior to a basic `nuts and bolts'-level description of how the spectral function can be modelled using this approach (Section 2.3). 
Finally in Section 2.4, a brief account is provided of how ARPES works so as to give experimental access to the spectral function.

Next - in Section 3 - we switch gears from the physics part of the paper to cover the philosophical aspects.
Here, we address four main conceptual questions concerned with how to interpret the successful semi-holographic description of these experimental data.


A \textbf{first foundational question} - covered in point (i) of Section 3.1 - is whether the success of semi-holography in describing the ARPES results from strange metals can be taken as empirical evidence for the relevance of holography to our real, physical world.
This is, naturally, connected to the issue of whether semi-holography is a distinct formal framework separate to holography, or whether it is - in fact - a type of holography.
Put another way, can semi-holography can be reduced to holography?
If the answer is `no' - as discussed in point (ii) of Section 3.1, no evidentiary connection can be made between the semi-holographic modelling of the experimental data and the significance of holography itself.
If the answer is `yes' - as discussed in point (iii) of Section 3.1- such a connection can be made.
This issue was discussed already in Section \ref{adscft}, and there the conclusion was reached that - at least for the semi-holographic models that form the focus of this paper - this reduction can indeed be done by taking an alternative, non-standard choice of boundary conditions in AdS-CFT, effectively taking the Legendre transform of AdS-CFT and integrating over the function that fixes the boundary condition. This ability to reduce semi-holography to holography then allows us to overcome several sceptical objections and clears the way for discussing the other philosophical questions.

The \textbf{second foundational question} in the discussion is presented in Section 3.2, and concerns the relation between semi-holography and emergence.
Three interesting options are identified, linked to the points discussed in Section 3.1, and classified in terms of three scenarios. 

The first - we call a \textit{vertical scenario} (A) - envisages emergence taking place in \textit{both} sides of the duality as illustrated in Figure \ref{Emergencefig1}. This scenario involves an affirmative answer to the question whether semi-holography is reducible to holography (point (iii) in Section 3.1). Here, the strange metal is the state that emerges, and this scenario can also result from a situation in which semi-holography and holography are both duals, yet each possessing a parameter (such as the operative energy scale) that can `tune' the emergent behaviour.     

The second - we call a \textit{horizontal scenario} (B) - is illustrated in Figure \ref{Emergencefig2} and relevant for the case where semi-holography is not a duality, i.e. it cannot be reduced to holography, as discussed in point (ii) of Section 3.1. In this case, the black hole emerges as the low-energy description of the strange metal.

The third - we call a \textit{diagonal scenario} (C) - sees the black hole emerge from the UV limit of the conformal field theory which can be visualized as being due to a combination of the renormalization group flow towards the IR in the field theory and the duality mapping of the IR strange metal onto the black hole.

The conclusion is that the black hole {\it can} indeed emerge from the field theory, even though we did not find epistemic justification for committing to the existence of the black hole and thus for being instrumentalists about the strange metal (we note that, as we discussed, if our black-hole monomaniac can find such epistemic justification, then they could have their cake and eat it.) As such, the conditions for emergence are less strict than those for explanation and scientific realism, because we are not here discussing metaphysical emergence and hence have more options.

The \textbf{third and fourth foundational questions} deal with explanation and scientific realism, respectively. These are related - as in both cases the interesting options are tied firstly to the possibility of reducing semi-holography to holography, and secondly to the existence of a common core between the dual theories.

The issue of explanation also lay behind the discussion in Section 3.1 (does holography explain the fact that experiments appear to provide empirical support for semi-holography?) and 3.2 (emergence as a tool for explanation).
In our discussion of explanation in Section 3.3, the focus was the possibility of using semi-holography to explain the properties of the strange metal, as opposed to merely being a neat, efficient computation device.
As it is very unlikely that a black hole is actually involved in the experiments being carried out, we conclude that the existence of a common core theory underlying the two duals is an appropriate stand-point, and that as discussed in Section \ref{adscft}, approaches to semi-holography that restore the duality between the bulk and the boundary are important in enabling such a common core theory to exist. 

Turning to scientific realism, the question is then posed whether the evidence from ARPES experiments in the form of the success of semi-holography to model the observed the k-dependent power laws describing the ($\omega,T$)-dependence of the strange metal's electron self-energy justifies the belief in the existence of the black hole. 
A trio of objections to this realist interpretation are provided and - on balance - we argue that although there is space for a consistent realist view, given the current information at our disposal this seems to involve arbitrary interpretative choices and is less convincing, leading to the general answer that - no - the experiments do not justify the assertion of the existence of the black hole.   

In the project described in this paper, condensed matter systems have been modelled using holography. This, then, ``inverts the explanatory arrows'' and this means we are taking the strange metal's anomalous behaviour as our explanandum (the thing that is to be explained).
As we have discussed above, the explanations are not given by the black hole as such, but rather by the common core theory that the black hole and the strange metal share.
This means that using the black hole and comparing the structure of the two theories will surely help to develop this common core.\footnote{It is beyond the scope of this paper to work out this common core. For gauge-gravity dualities, the common core theory is discussed in De Haro (2020). For the common core of T-duality, see Cinti and De Haro (2025). For other examples and further discussion of common core theories, see De Haro and Butterfield (2025): especially Sections 4.1, 5.1.2, 5.3, 12.2, and 12.4. Section 13.2 discusses the sense in which a commitment to the common core is epistemically justified.}
In this sense, rather than a realist interpretation focused on the existence of the black hole, there can then be a justified realist commitment to the common core theory. This epistemically justified commitment to the common core is then - in turn - a commitment to holography itself, in keeping with the empirical evidence coming from the ARPES experiments.

\section*{Acknowledgements}
\addcontentsline{toc}{section}{Acknowledgements}
\markboth{\small{\textup{Acknowledgements}}}{\textup{\small{Acknowledgements}}}

MSG, SDH, and EC acknowledge the funding they have received from the Dutch National Science Agenda (Nationale Wetenschapsagenda; Small Innovative NWA projects) by NWO, under project number NWA.1418.22.029 and title {\it Is There Space and Time for Experimental Philosophy?}

\section*{References}
\addcontentsline{toc}{section}{References}
\markboth{\small{\textup{References}}}{\textup{\small{References}}}

Butterfield, J.~N.~(2011a). `Emergence, Reduction and Supervenience: A Varied Landscape'. {\it Foundations of Physics}, 41, pp.~920-959.

\ \\Butterfield, J.~N.~(2011b). `Less is Different: emergence and Reduction Reconciled'. {\it Foundations of Physics}, 41, pp.~1065-1135.

\ \\Butterfield, J.~N~(2021). `On Dualities and Equivalences Between Physical Theories'. In: {\it Philosophy Beyond Spacetime}, Huggett, N., Le Bihan B.~and W\"uthrich, C.~(Eds.), pp.~41-77. Oxford: Oxford University Press.

\ \\Casalderrey-Solana, J., Liu, H., Mateos, D., Rajagopal, K., Achim Wiedemann, U. (2014). {\it Gauge/string duality, hot QCD and heavy ion collisions}. Cambridge University Press.

\ \\Chakraborty, T., Chakravarty, J., Godet, V., Paul, P., and Raju, S.~(2023). `Holography of information in de Sitter space'. {\it Journal of High-Energy Physics}, 2023(12), pp.~1-46.

\ \\Chakravartty, A.~(2017). `Scientific Realism'. {\it Stanford Encyclopedia of Philosophy}. https://plato.stanford.edu/entries/scientific-realism.

\ \\Chandrasekaran, V., Longo, R., Penington, G., and Witten, E.~(2023). `An algebra of observables for de Sitter space'. {\it Journal of High-Energy Physics}, 2023(2), pp.~1-56.

\ \\ Contino, R.,  Pomarol, A. (2004). `Holography for fermions'. {\it Journal of High-Energy Physics}, 2004(11), 058.

\ \\Crowther, K.~(2016). {\it Effective Spacetime. Understanding Emergence in Effective Field Theory and Quantum Gravity}. Springer.

\ \\Curd, M., and Cover, J. A.~(1998). `Philosophy of Science: The Central Issues'. {\it New York and London: WW Norton and Company}.

\ \\Dardashti, R., Dawid, R., Gryb, S., and Thébault, K.~(2018). `On the Empirical Consequences of the AdS/CFT Duality'. In: {\it Beyond Spacetime: The Foundations of Quantum Gravity}, Huggett, N., Matsubara, K.,~and W\"uthrich, C.~(Eds.), pp. 284-303. Cambridge University Press.

\ \\Dawid, R.~(2017). `String dualities and empirical equivalence'. {\it Studies in History and Philosophy of Science Part B: Studies in History and Philosophy of Modern Physics}, 59, pp.21-29.

\ \\De Haro, S.~(2017). `Dualities and emergent gravity: Gauge/gravity duality'. {\it Studies in History and Philosophy of Modern Physics}, 59, pp.~109-125. 

\ \\De Haro, S.~(2019). `Towards a Theory of Emergence for the Physical Sciences'. {\it European Journal for Philosophy of Science}, 9 (38), pp.~1-52.

\ \\De Haro, S.~(2020). `Spacetime and Physical Equivalence'. In: {\it Beyond Spacetime. The Foundations of Quantum Gravity}, Huggett, N., Matsubara, K.~and W\"uthrich, C.~(Eds.), pp.~257-283. Cambridge: Cambridge University Press. 

\ \\De Haro, S.~(2023). `The empirical under-determination argument against scientific realism for dual theories'. {\it Erkenntnis}, 88(1), pp.~117-145.

\ \\De Haro, S.~and Butterfield, J.~N.~(2025). {\it The Philosophy and Physics of Duality}, forthcoming at Oxford University Press.

\ \\De Haro, S., Mayerson, D.~R.~and Butterfield, J.~N.~(2016). `Conceptual aspects of gauge/gravity duality'. {\it Foundations of Physics}, 46, pp.~1381-1425.

\ \\Dieks, D., van Dongen, J.~and De Haro, S.~(2015). `Emergence in Holographic Scenarios for Gravity'. {\it Studies in History and Philosophy of Modern Physics}, 52, pp.~203-216.

\ \\Eskens, F.~D.~J.~(2025). `On effective dualities'. {\it Synthese}, 206 (3), 155.

\ \\ Faulkner, T., Polchinski, J. (2011). `Semi-holographic Fermi liquids'. {\it Journal of High-Energy Physics}, 2011(6), 1-23.

\ \\Guay, A., Sartenaer, O.~(2016). `A new look at emergence. Or when after is different'. {\it European Journal for Philosophy of Science}, 6 (2), pp.~297-322.

\ \\ Gubser, S. S., Klebanov, I. R., Polyakov, A. M. (1998). `Gauge theory correlators from non-critical string theory'. {\it Physics Letters B}, 428(1-2), 105-114.

\ \\ Gubser, S. S. and Rocha, F. D. (2010). `Peculiar properties of a charged dilatonic black hole in AdS5'. {\it Physical Review D}, 81, 1335.

\ \\G\"ursoy, U.~(2021). `Holographic QCD and magnetic fields'. {\it The European Physical Journal A}, 57, 247.

\ \\G\"ursoy, U., Plauschinn, E., Stoof, H.~and Vandoren, S.~(2012). `Holography and ARPES sum-rules'. {\it Journal of High-Energy Physics}, 2012(5), pp.~1-21.

\ \\Hartnoll, S. A., Lucas, A.,  Sachdev, S. (2018). Holographic quantum matter. MIT press.

\ \\Humphreys, P.~(2016). {\it Emergence. A Philosophical Accounts}. New York and Oxford: Oxford University Press.

\ \\ Kovtun, P. K., Son, D. T.,  Starinets, A. O. (2005). `Viscosity in strongly interacting quantum field theories from black-hole physics'. {\it Physical review letters}, 94(11), 111601.

\ \\Le Bihan, B.~and Read, J.~(2018). `Duality and Ontology'. Published online in {\it Philosophy Compass}.

\ \\ Maldacena, J. (1999). `The large-N limit of superconformal field theories and supergravity.' {\it International journal of theoretical physics}, 38(4), 1113-1133.

\ \\Matsubara, K.~(2013). `Realism, underdetermination and string theory dualities'. {\it Synthese}, 190 (3), pp.~471-489.

\ \\Mauri, E., Smit, S., Golden, M. S., \& Stoof, H. T. C.~(2024). `Gauge-gravity duality comes to the laboratory: Evidence of momentum-dependent scaling exponents in the nodal electron self-energy of cuprate strange metals'. {\it Physical Review B}, 109(15), 155140.

\ \\M\o ller-Nielsen, T.~(2017). `Invariance, Interpretation, and Motivation'. {\it Philosophy of Science}, 84 (5), pp.~1253-1264.

\ \\North, J.~(2021). {\it Physics, Structure, and Reality}. Oxford: Oxford University Press. 

\ \\ Faulkner, T.~and Polchinski, J.~(2011). `Semi-holographic Fermi liquids'. {\it Journal of High-Energy Physics}, 2011(6), 1-23.

\ \\ Policastro, G., Son, D. T., Starinets, A. O. (2001). `Shear viscosity of strongly coupled N= 4 supersymmetric Yang-Mills plasma'. {\it Physical Review Letters}, 87(8), 081601.

\ \\Rickles, D.~(2011). `A Philosopher looks at String Dualities'. {\it Studies in History and Philosophy of Modern Physics}, 42 pp.~54-67.

\ \\Smit, S. et al., ~(2024). `Momentum-dependent  scaling exponents of nodal self energies measured in strange metal cuprates and modelled using semi-holography'. {\it Nature Communications
}, 15, 4581.

\ \\Strominger, A.~(2001). `The dS/CFT correspondence'. {\it Journal of High-Energy Physics}, 2001(10), 034.

\ \\Teh, N. J.~(2013). `Holography and emergence'. {\it Studies in History and Philosophy of Science Part B: Studies in History and Philosophy of Modern Physics}, 44(3), pp.300-311.

\ \\Williams, P.~(2019). `Scientific Realism Made Effective'. {\it The British Journal for the Philosophy of Science}, 70, pp.~209-237.

\ \\ Witten, E. (1998). 'Anti de Sitter space and holography'. {\it Advances in Theoretical and Mathematical Physics} 2 (1998) 253-291, arXiv preprint hep-th/9802150.

\ \\Witten, E.~(2001). `Quantum gravity in de Sitter space'. arXiv preprint hep-th/0106109.

\ \\Woodward, J. and Ross, L.~(2021), `Scientific Explanation', {\it The Stanford Encyclopedia of Philosophy}, https://plato.stanford.edu/archives/sum2021/entries/scientific-explanation.

\ \\Zaanen, J., Liu, Y., Sun, Y. W.~and Schalm, K. (2015). {\it Holographic duality in condensed matter physics}. Cambridge University Press.

\ \\Zhang, H., Pincelli, T., Jozwiak, C., Kondo, T., Ernstorfer, R., Sato, T., \& Zhou, S.~(2022). `Angle-resolved photoemission spectroscopy'. {\it Nature Reviews Methods Primers}, 2(1), 54.


\end{document}